\documentclass[12pt]{article}
\usepackage{amsfonts}
\usepackage{amssymb}
\usepackage{graphics,psboxit,amsmath}
\usepackage{subfigure}
\usepackage{graphicx}
\usepackage{verbatim}
\usepackage{epic}

\def\hybrid{\topmargin 0pt \oddsidemargin 0pt 
        \headheight 0pt \headsep 0pt
        \textwidth 16,5cm 
        \textheight 23,5cm 
        \marginparwidth .875in
        \parskip 5pt plus 1pt \jot = 1.5ex}

\hybrid

\def\baselinestretch{1.2}

\catcode`\@=11

\def\marginnote#1{}
\newcount\hour
\newcount\minute
\newtoks\amorpm
\hour=\time\divide\hour by60
\minute=\time{\multiply\hour by60 \global\advance\minute by-\hour}
\edef\standardtime{{\ifnum\hour<12 \global\amorpm={am}%
        \else\global\amorpm={pm}\advance\hour by-12 \fi
        \ifnum\hour=0 \hour=12 \fi
        \number\hour:\ifnum\minute<10 0\fi\number\minute\the\amorpm}}
\edef\militarytime{\number\hour:\ifnum\minute<10 0\fi\number\minute}

\def\draftlabel#1{{\@bsphack\if@filesw {\let\thepage\relax
   \xdef\@gtempa{\write\@auxout{\string
      \newlabel{#1}{{\@currentlabel}{\thepage}}}}}\@gtempa
   \if@nobreak \ifvmode\nobreak\fi\fi\fi\@esphack}
        \gdef\@eqnlabel{#1}}
\def\@eqnlabel{}
\def\@vacuum{}
\def\draftmarginnote#1{\marginpar{\raggedright\scriptsize\tt#1}}

\def\draft{\oddsidemargin -.5truein
        \def\@oddfoot{\sl preliminary draft \hfil
        \rm\thepage\hfil\sl\today\quad\militarytime}
        \let\@evenfoot\@oddfoot \overfullrule 3pt
        \let\label=\draftlabel
        \let\marginnote=\draftmarginnote
   \def\@eqnnum{(\theequation)\rlap{\kern\marginparsep\tt\@eqnlabel}%
\global\let\@eqnlabel\@vacuum} }

\def\draft2{
        \def\@oddfoot{\sl preliminary draft \hfil
        \rm\thepage\hfil\sl\today\quad\militarytime}
        \let\@evenfoot\@oddfoot \overfullrule 3pt
        \let\label=\draftlabel
        \let\marginnote=\draftmarginnote
   \def\@eqnnum{(\theequation)\rlap{\kern\marginparsep\tt\@eqnlabel}%
\global\let\@eqnlabel\@vacuum} }

\def\preprint{\twocolumn\sloppy\flushbottom\parindent 2em
        \leftmargini 2em\leftmarginv .5em\leftmarginvi .5em
        \oddsidemargin -.5in \evensidemargin -.5in
        \columnsep .4in \footheight 0pt
        \textwidth 10.in \topmargin -.4in
        \headheight 12pt \topskip .4in
        \textheight 6.9in \footskip 0pt
        \def\@oddhead{\thepage\hfil\addtocounter{page}{1}\thepage}
        \let\@evenhead\@oddhead \def\@oddfoot{} \def\@evenfoot{} }

\def\numberbysection{\@addtoreset{equation}{section}
        \def\theequation{\thesection.\arabic{equation}}}

\def\underline#1{\relax\ifmmode\@@underline#1\else
        $\@@underline{\hbox{#1}}$\relax\fi}

\def\titlepage{\@restonecolfalse\if@twocolumn\@restonecoltrue\onecolumn
     \else \newpage \fi \thispagestyle{empty}\c@page\z@
        \def\thefootnote{\fnsymbol{footnote}} }

\def\endtitlepage{\if@restonecol\twocolumn \else \newpage \fi
        \def\thefootnote{\arabic{footnote}}
        \setcounter{footnote}{0}} 

\catcode`@=12
\relax

\def\figcap{\section*{Figure Captions\markboth
        {FIGURECAPTIONS}{FIGURECAPTIONS}}\list
        {Figure \arabic{enumi}:\hfill}{\settowidth\labelwidth{Figure
999:}
        \leftmargin\labelwidth
        \advance\leftmargin\labelsep\usecounter{enumi}}}
 \relax
\def\tablecap{\section*{Table Captions\markboth
        {TABLECAPTIONS}{TABLECAPTIONS}}\list
        {Table \arabic{enumi}:\hfill}{\settowidth\labelwidth{Table
999:}
        \leftmargin\labelwidth
        \advance\leftmargin\labelsep\usecounter{enumi}}}
 \relax
\def\reflist{\section*{References\markboth
        {REFLIST}{REFLIST}}\list
        {[\arabic{enumi}]\hfill}{\settowidth\labelwidth{[999]}
        \leftmargin\labelwidth
        \advance\leftmargin\labelsep\usecounter{enumi}}}
 \relax
\makeatletter
\newcounter{pubctr}
\def\publist{\@ifnextchar[{\@publist}{\@@publist}}
\def\@publist[#1]{\list
        {[\arabic{pubctr}]\hfill}{\settowidth\labelwidth{[999]}
        \leftmargin\labelwidth
        \advance\leftmargin\labelsep
        \@nmbrlisttrue\def\@listctr{pubctr}
        \setcounter{pubctr}{#1}\addtocounter{pubctr}{-1}}}
\def\@@publist{\list
        {[\arabic{pubctr}]\hfill}{\settowidth\labelwidth{[999]}
        \leftmargin\labelwidth
        \advance\leftmargin\labelsep
        \@nmbrlisttrue\def\@listctr{pubctr}}}
 \relax
\makeatother

\def\ba{\begin{equation}}
\def\ea{\end{equation}}

\def\del{\partial}

\def\no{\noindent}

\def\IR{\relax{\rm I\kern-.18em R}}

\begin{document}


\renewcommand{\theequation}{\thesection.\arabic{equation}}
\csname @addtoreset\endcsname{equation}{section}

\newcommand{\eqn}[1]{(\ref{#1})}
\newcommand{\be}{\begin{eqnarray}}
\newcommand{\ee}{\end{eqnarray}}
\newcommand{\non}{\nonumber}

\begin{titlepage}
\strut\hfill
\vskip 1.3cm
\begin{center}

{\large \bf Liouville integrable defects: the non-linear Schr\"{o}dinger paradigm}

\vskip 0.5in

{\bf Jean Avan$^{a}$ and  Anastasia Doikou$^{b}$}
\\[8mm]
\noindent
{\footnotesize  $^a$ LPTM, Universite de Cergy-Pontoise (CNRS UMR 8089),
F-95302 Cergy-Pontoise, France}
\\
{\footnotesize {\tt E-mail: avan@u-cergy.fr}}
\\[4mm]
\noindent
{\footnotesize $^b$
Department of Engineering Sciences, University of Patras,
GR-26500 Patras, Greece}
\\
{\footnotesize {\tt E-mail: adoikou@upatras.gr}}

\end{center}

\vskip .6in

\centerline{\bf Abstract}
A systematic approach to Liouville integrable defects is proposed, based on an underlying Poisson algebraic structure.
The non-linear Schr\"{o}dinger model in the presence of a single particle-like defect is investigated through this algebraic approach.
Local integrals of motions are constructed as well as the time components of the corresponding Lax pairs.
Continuity conditions imposed upon the time components of the Lax pair to all orders give rise to sewing conditions,
which turn out to be compatible with the hierarchy of charges in involution. Coincidence of our results with the continuum limit of
the discrete expressions obtained in earlier works further confirms our approach.

\no

\vfill
\no

\end{titlepage}
\vfill

\eject


\tableofcontents

\def\baselinestretch{1.2}
\baselineskip 20 pt
\no

\section{Introduction}

The issue of integrable defects in discrete and continuum  (classical-quantum) integrable systems has been the subject of increased research interest during the last two decades or so \cite{delmusi}--\cite{weston}.
Recently one of us \cite{doikou-defect} proposed an algebraic approach for the description of a Liouville integrable
defect in the discrete non-linear Schr\"{o}dinger model. This approach is based on the construction of a $N$-site transfer matrix including the
defect matrix at a fixed point. Classical integrability is guaranteed by the existence of an $r$-matrix structure \cite{sts} for the
discrete bulk Lax matrices, and the defect matrix. The question of quantum integrable defects (see e.g. \cite{haku, weston} and references therein) will not be treated here, although the corresponding formalism is a straightforward variation of the classical one \cite{doikou-defect}.
Subsequent derivations of the Poisson-commuting Hamiltonians
and the corresponding time-component of the Lax pair following the canonical construction \cite{sts, FT} were given.

This now leads us to propose a similarly fully algebraic picture for a description of a Liouville integrable defect
in the continuous non-linear Schr\"{o}dinger model. We restrict ourselves to the case of a single point like defect;
extensions of this notion will be commented upon in the conclusion section.

The procedure itself is based on the construction of a suitable continuous transfer matrix
generating the Poisson-commuting Hamiltonians and their associated time-component ${\mathbb V}$ of the continuous Lax pair:
\\
\\
{\bf 1.} The continuous monodromy matrix is built as a coaction:
\be
T(A, -A, \lambda) = T^+(A, x_0, \lambda)\ \tilde  L(x_0, \lambda)\ T^-(x_0, -A, \lambda) \label{monodefect}
\ee
This of course is the immediate continuum limit of the discrete defect monodromy matrix
(see e.g. \cite{doikou-defect, ADS}). Such monodromy matrices were derived in \cite{maillet}. The $T^{\pm}$ matrices are the
monodromies of the differential operator $d/dx + L(x)$ where $L$ is the continuous Lax matrix $L(x)$ associated to NLS \cite{kundura},
and $\tilde L$ is the defect matrix. As in the discrete case, Liouville integrability follows from asking that $\tilde L$ obeys a
quadratic Poisson algebra
\be
\Big \{ \tilde L_a(\lambda),\ \tilde L_b(\mu)\Big \} = \Big [ r_{ab}(\lambda-\mu),\ \tilde  L_a(\lambda) \tilde  L_b(\mu)\Big ] \label{rtt}
\ee
with the same $r$ matrix as the bulk monodromy operators, thereby imposing a strong constraint on the Poisson structure
of the dynamical variables parametrizing the defect.
\\
\\
{\bf 2.} The Poisson-commuting hierarchy of Hamiltonians is then obtained from expansion in $\lambda^{-1}$ of the
$ln$ of the trace of the monodromy matrix (\ref{monodefect}). Poisson commutation is formally guaranteed by the underlying
quadratic Poisson structure \cite{FT}, but must be checked against possible divergences due to $\delta$-distributions
on a support overlapping the defect point.
\\
\\
{\bf 3.} The time components of the Lax pair are then computed. They are evaluated separately in the right
bulk $(x_0,\ A)$ and the left bulk  $(-A,\ x_0)$ and on the defect point --from left and right. As
in the boundary integrable systems \cite{avandoikou} it is required that ${\mathbb V}^{(\pm)}(x_0^{\pm}) \to \tilde {\mathbb V}^{(\pm)}(x_0)$
in order to avoid singular contributions from the zero curvature condition for the Lax pair ${\mathbb U},\ {\mathbb V}$:
\be
\dot{{\mathbb U}}- {\mathbb V}' + \Big [{\mathbb U},\ {\mathbb V} \Big ]=0, ~~~~x\neq x_0. \label{zero0}
\ee
This translates into sewing conditions $\{C^{(j)}_{\pm} \}$ across the defect relating the right and left values of the
$(j-1)$th derivatives of the fields by functions of lower derivatives and the defect parameters. Sewing conditions
 are thus understood as necessary conditions to allow identification of the Hamiltonian equations of motion deduced either
from $H^{(i)}$ or from the zero curvature condition for the Lax pair ${\mathbb U},\ {\mathbb V}^{(i)}$; in other words they
act as ``regularizations'' in the canonical \cite{sts, FT} procedure yielding ${\mathbb V}^{(i)}$ and $H^{(i)}$ through the
classical $r$-matrix. They will be shown in our example to be {\it sufficient} conditions.
\\
\\
{\bf 4.} Consistency of the procedure then requires to make sure that the sub-manifold of the sewing conditions
$\{C_{\pm}^{(i)}\}$ is invariant under the Hamiltonian action, which reads as:
\be
\Big \{ {\cal H}^{(i)},\ C_{\pm}^{(j)} \Big \} \ \mbox{belongs to the ideal generated by $C^{(i)}_{\pm}$}. \label{sew}
\ee
Once this is checked, we are justified in defining our Hamiltonian dynamical system as a Liouville-integrable defect in the continuum.

As it clearly appears from our construction in Point 1, the monodromy matrix with defect realizes a Hamiltonian formulation for a Backlund
(or rather dressing) transformation procedure yielding an ``integrable'' defect in a Lagrangian approach such as
described previously in \cite{BCZ1} and recently in a very explicit way in \cite{newdefect}. Because we are from the very beginning
in a Hamiltonian framework involving the $r$-matrix structure and associated construction of the Lax pair, we are indeed
safe in stating that we are establishing  a strong Liouville-integrability for our defect theory. In addition this provides
the suitable basis for a quantization procedure. By contrast integrability discussed in
\cite{BCZ1, newdefect} can be characterized as weaker ``Lax''-type integrability, in that they only
show the existence of modified conserved quantities and their invariance under time evolution triggered by
(in the Hamiltonian language) the third Hamiltonian. Higher time evolution cannot be discussed in this Lagrangian
framework hence Liouville integrability can not be proved.

Exemplification of the fourfold pattern will be given in Section $2$ on the example of NLS equation.
It must be emphasized that our direct construction can also be formulated as a continuous
limit of the discrete construction given in \cite{doikou-defect}. We shall comment on this in a final section $3$: The continuous limit is here formulated as the replacement of the discrete
index $n$ by the continuous variable $x$; the introduction of the  normalization
scale $\Delta$ in the bulk matrices exactly as explained in Section $5$ of the previous paper;
and the introduction of this scale $\Delta$ and an overall factor on the defect
Lax matrix (disregarding here the parametrization used in
\cite{doikou-defect}, which might be relevant in the case of non ultra-local or extended defects).

Finally, a general proof of the
consistency of such sewing conditions as obtained in Step $3$ with the Hamiltonian evolutions triggered by a Lax pair
formulation will be given in Section $4$.

The Hamiltonians obtained by this procedure; the Lax time-operators ${\mathbb V}(x)$;
and the sewing conditions are exactly identified with the direct continuous
construction described in the previous sections. It appears that the  ``naive'' discrete
to continuous limit is here consistent, which may be related to the
ultra-local nature of the considered theory.

\section{The continuous NLS model with defect}

The four fold pattern described in the introduction is generic. We shall now exemplify it on the simple example of a single
point-like defect in a  continuum field theory --the non-linear Schr\"{o}dinger model-- associated to the Yangian classical
$r$-matrix \cite{yang}: $\ r(\lambda) = {{\cal P} \over \lambda}$, $\ {\cal P}$ is the permutation operator.

The starting point in our analysis is the derivation of the corresponding monodromy matrix:
\be
T(A, -A, \lambda)&=& T^+(A, x_0, \lambda)\ \tilde L(x_0, \lambda)\ T^-(x_0, -A,\lambda)\non\\ &=& P\exp \Big \{\int_{x_0^+}^A dx\ {\mathbb U}^+(x) \Big \}\ \tilde L(x_0,\lambda)\ P\exp\Big \{\int_{-A}^{x_0^-}dx\ {\mathbb U}^-(x) \Big \} \label{contmon}
\ee
$T^{\pm}$ and $\tilde L$ satisfy the quadratic relation (\ref{rtt}).
We consider the following defect operator
\be
\tilde L(x_0) = \lambda {\mathbb I} + \begin{pmatrix}
\alpha(x_0) &\beta(x_0)\\
\gamma(x_0) &\delta(x_0)
\end{pmatrix}.
\ee
The Lax operator for the NLS model is the familiar (see e.g. \cite{FT, doikouit}):
\be
\mathbb{U}^{\pm} = \mathbb{U}_d + \mathbb{U}^+_a \equiv \frac{\lambda}{2}
\begin{pmatrix}
  1 & 0 \cr
  0 & -1
\end{pmatrix}
+
\begin{pmatrix}
  0 & \bar \psi^{\pm} \cr
  \psi^{\pm} & 0
\end{pmatrix}.
\ee
where the bulk fields are canonical i.e.
\be
\Big \{ \psi^{\pm}(x),\ \bar \psi^{\pm}(y) \Big \} = \delta(x-y), ~~~~~\Big \{ \psi^{\mp}(x),\ \bar \psi^{\pm}(y) \Big \}=0. \label{ex1}
\ee
Due to the fact that the $\tilde L$ satisfies the quadratic algebra (\ref{rtt})
the elements $\alpha,\ \beta,\ \gamma,\ \delta$ realize the following Poison bracket structure:
\be
&& \Big \{\alpha(x_0),\ \beta(x_0)\Big \}= \beta(x_0) \non\\
&& \Big \{\alpha(x_0),\ \gamma(x_0)\Big \} = -\gamma(x_0) \non\\
&& \Big \{\beta(x_0),\ \gamma(x_0)\Big \} = 2 \alpha(x_0). \label{ex2}
\ee
The discussion on the continuum limit of the discrete NLS in the section 3 will further justify the present analysis.
It will become transparent that the results derived directly from the continuum monodromy matrix coincide,
as one would naturally expect, with the ones obtained as continuum limits of the discrete expressions presented in
section $3$ . In particular, it will be clear (see also \cite{ADS, doikoucl} and expression (\ref{contb})) that the
continuum analogue of the discrete monodromy matrix is given by (\ref{contmon}).

The continuum ``bulk'' monodromy matrices $T^{\pm}$ satisfy the following differential equation
\be
{\partial T^{\pm}(x,y; \lambda) \over \partial x} = {\mathbb U}^{\pm} T^{\pm}(x,y; \lambda) \label{dif1}
\ee
and the zero curvature condition is then expressed as:
\be
\dot{\mathbb U}^{\pm}(x,t) - {\mathbb V}^{\pm '}(x,t) + \Big [{\mathbb U}^{\pm}(x, t), {\mathbb V}^{\pm}(x, t) \Big ] =0 ~~~~~x \neq x_0
\label{zero1}\ee
On the defect point in particular the zero curvature condition is formulated as (this will be also
transparent when discussing the continuum limit of the discrete theory)
\be
{d \tilde L(x_0) \over dt}  = \tilde {\mathbb V}^+(x_0) \tilde L(x_0) - \tilde L(x_0) \tilde {\mathbb V}^{-}(x_0) \label{zerod}
\ee
and describes explicitly the jump occurring across the defect point. This will be a
major consistency check of the prescription followed here. The time components $\tilde {\mathbb V}^{\pm}$ at the defect point
will be explicitly derived below together with the bulk quantities ${\mathbb V}^{\pm}$ and the ``defect'' quantities $\tilde {\mathbb V}^{\pm}$.

Now the canonical procedure in extracting the local integrals of motion may be directly applied. First we
consider the following familiar ansatz for the monodromy matrices:
\be
T^{\pm}(x,y;\lambda) = (1+W^{\pm}(x))e^{Z^{\pm}(x,y)}(1+W^{\pm}(y))^{-1} \label{transatz}
\ee

Substituting the ansatz (\ref{transatz}) into \eqn{dif1}, and splitting the
resulting equation into a diagonal and an off-diagonal part one obtains
\be
&& \frac{d W^{\pm}}{dx} + W^{\pm} \mathbb{U}_d - \mathbb{U}_d W^{\pm} +  W^{\pm} \mathbb{U}_a^{\pm} W^{\pm}- \mathbb{U}^{\pm}_a =0,\cr
&& \frac{\del Z^{\pm}}{\del x} = \mathbb{U}_d + \mathbb{U}_a^{\pm} W^{\pm}. \label{split0}
\ee
Solution of the latter set of equations provides the explicit expressions of the $W^{\pm},\ Z^{\pm}$ matrices.

More precisely, let us recall the generating function of the local integrals of motion
\be
{\cal G}(\lambda)=  \ln \Big (tr T(\lambda)\Big )
\ee
due to the ansatz (\ref{transatz}) we can substitute the monodromy matrix accordingly and obtain:
\be
{\cal G}(\lambda)= \ln tr \Big [(1+W^+(L)) e^{Z^+(A, x_0)}(1+W^+(x_0))^{-1}
\tilde L(x_0) (1+W^-(x_0)) e^{Z^-(x_0, -A))} (1 +W^-(-L))^{-1}\Big ], \non\\
\ee
but due to the choice of Schwartz boundary conditions at $x = \pm A$
we conclude:
\be
{\cal G}(\lambda) = \ln tr \Big [e^{Z^+(A, x_0)} (1+W^+(x_0))^{-1} \tilde L(x_0) (1+W^-(x_0)) e^{Z^-(x_0, -A))}  \Big ].
\ee
Let us first evaluate the first couple of $W^{(i)}$'s through (\ref{split0})
\be
&& W^{\pm(1)} = \begin{pmatrix}
    &       -\bar \psi^{\pm}(x) \\
    \psi^{\pm}(x) &
\end{pmatrix} \ \label{1lmatrix}, ~~~~~W^{\pm(2)} = \begin{pmatrix}
    &       -\bar \psi^{\pm'}(x) \\
    -\psi^{\pm'}(x) &
\end{pmatrix} \ \label{2lmatrix} \non\\
&& W^{\pm(3)} = \begin{pmatrix}
    &       -\bar \psi^{\pm ''}(x) + |\psi^{\pm}(x)|^2 \bar \psi^{\pm}(x)\\
    \psi^{\pm''}(x)- |\psi^{\pm}(x)|^2  \psi^{\pm}(x)&
\end{pmatrix}. \label{3lmatrix}
\ee

Similarly through (\ref{split0}) the diagonal elements $Z^{(i)}$ are given as:
\be
&& Z^{+(-1)} = {1\over 2}\begin{pmatrix}
  A-x_0  &        \\
     & -A+x_0
\end{pmatrix} \  ,~~~~ Z^{-(-1)} ={1\over 2} \begin{pmatrix}
  A+x_0  &        \\
     & -A-x_0
\end{pmatrix} \   \label{4lmatrix},\non\\
&& Z^{\pm(1)} = \begin{pmatrix}
   \int dx\ \psi^{\pm}(x) \bar \psi^{\pm}(x)  &       \\
  &-\int dx\ \psi^{\pm}(x) \bar \psi^{\pm}(x)
\end{pmatrix} \ \label{5lmatrix}\non\\
&& Z^{\pm(2)} = \begin{pmatrix}
  -\int dx\ \psi^{\pm'}(x) \bar \psi^{\pm}(x)  &     \\
     & -\int dx\ \psi^{\pm}(x) \bar \psi^{\pm'}(x)
\end{pmatrix} \ \label{6lmatrix} \non\\
&& Z^{\pm(3)} = \begin{pmatrix}
  \int dx\ \Big ( \psi^{\pm''}(x) \bar \psi^{\pm}(x) - |\psi^{\pm}(x)|^4  \Big )  &       \\
            & -\int dx\ \Big (\bar \psi^{\pm''}(x) \psi^{\pm}(x)- |\psi^{\pm}(x)|^4  \Big )
\end{pmatrix}. \label{7lmatrix}
\ee
The expansion of the generating function ${\cal G}$ in powers of ${1\over \lambda}$ provides
the local integrals of motion of the model under consideration. Note that due to the fact that for $\lambda \to \infty$
the leading contribution comes from $Z^{(-1)}_{11}$ --keep also in mind that $A \to \infty$.
It is thus clear that the generating function of the local integrals of motion becomes:
\be
{\cal G}(\lambda) =Z_{11}^{+}(\lambda) + Z^{-}_{11}(\lambda) + \ln [(1+W^+(x_0))^{-1}\tilde L(x_0)(1+W^-(x_0))]_{11}
\ee
and the terms $Z_{11}^{\pm}$ provide the left and right bulk charges, whereas the third term of
the expression above gives the defect contribution.
More precisely, the first three integrals of motions may be expressed as
\be
{\cal H}^{(1)} = \int_{-A}^{x_0^-} dx\ \psi^{-}(x) \bar \psi^-(x) + \int^{A}_{x_0^+} dx\ \psi^{+}(x) \bar \psi^+(x) +\alpha(x_0). \label{h1}
\ee
\be
{\cal H}^{(2)} &=&  -\int_{-A}^{x_0^-} dx\ \bar \psi^{-}(x) \psi^{-'}(x) -\int^{A}_{x_0^+}dx\ \bar \psi^{+}(x)  \psi^{+'}(x) \non\\ & -&  \bar \psi^+(x_0)  \psi^+(x_0) + \bar  \psi^+(x_0)  \psi^-(x_0) + \gamma(x_0) \bar  \psi^+(x_0) +\beta(x_0)  \psi^-(x_0) -{\alpha^2(x_0)\over 2}  \non\\ \label{h2}
\ee
\be
{\cal H}^{(3)} &=& \int_{x_0^+}^A dx\ \Big ( \bar  \psi^+(x) \psi^{+''}(x) + |\psi^+(x)|^4\Big ) + \int_{-A}^{x_0^-}dx\ \Big (\bar  \psi^-(x) \psi^{-''}(x) + |\psi^-(x)|^4 \Big) \non\\ &+& (\bar  \psi^+(x_0) \psi^+(x_0))' + \gamma(x_0)\bar  \psi^{+'}(x_0) - \beta(x_0) \psi^{-'}(x_0) +\bar \psi^{+'}(x_0)\psi^-(x_0) +{\alpha^3(x_0)\over 3} \non\\
&-& \bar  \psi^{+}(x_0)  \psi^{-'}(x_0) -\alpha(x_0)\Big (\gamma(x_0)\bar  \psi^+(x_0) +\beta(x_0) \psi^-(x_0)+2 \bar \psi^{+}(x_0) \psi^-(x_0) \Big ). \non\\\label{h3}
\ee
It is clear that by construction for all the above charges we have:
\be
\Big \{ {\cal H}^{(i)},\ {\cal H}^{(j)} \Big\} =0,
\ee
the latter commutation relations have also been explicitly checked for the three charges (\ref{h3}).
To show the commutativity of the charges we made use of the exchange relations (\ref{ex1}), (\ref{ex2}) and also
\be
\Big \{\psi^{\pm}(x_0), \ {\mathrm e}(x_0) \Big \} =0, ~~~~{\mathrm e} = \Big \{ \alpha, \ \beta,\ \gamma \Big \}.
\ee
The fields $\psi^{\pm},\ \bar \psi^{\pm}$ at the defect are defined point by analytic continuation:
\be
\psi^{\pm}(x_0^{\pm}) \to \psi^{\pm}(x_0), ~~~~~~\bar \psi^{\pm}(x_0^{\pm}) \to \bar \psi^{\pm}(x_0).
\ee

Expressions of the time component ${\mathbb V}$ of the Lax pair are known (see e.g. \cite{FT}).
The generic expressions for the bulk left and right theory are given as:
\be
&&{\mathbb V}^{+}(x, \lambda, \mu) = t^{-1}(\lambda) tr_a \Big (T_a^+(A, x ,\lambda) r_{ab}(\lambda -\mu)T_a^+(x, x_0, \lambda)
 \tilde L_a(x_0, \lambda)T_a^-(x_0, -A, \lambda) \Big ) \non\\
&& {\mathbb V}^{-}(x, \lambda, \mu) = t^{-1}(\lambda) tr_a \Big (T_a^+(A, x_0 ,\lambda)\tilde L_a(x_0, \lambda) T_a^-(x_0,x,\lambda)
r_{ab}(\lambda -\mu)T_a^-(x, -A, \lambda) \Big ) \non\\
&& \tilde {\mathbb V}^+(x_0, \lambda, \mu) = t^{-1}(\lambda) tr_a \Big ( T_a^+(A, x_0, \lambda) r_{ab}(\lambda-\mu)
\tilde L_a(x_0, \lambda) T_a^-(x_0,-A, \lambda)\Big )\non\\
&& \tilde {\mathbb V}^-(x_0,\lambda, \mu)= t^{-1}(\lambda) tr_a \Big (T_a^+(A, x_0, \lambda) \tilde L_a(x_0, \lambda)
r_{ab}(\lambda -\mu)T_a^-(x_0, -A, \lambda) \Big ).
\label{timecomp}
\ee

In the special case where the $r$ matrix is the Yangian solution the expressions above become:
\be
&&{\mathbb V}^+(x, \lambda, \mu) = {t^{-1} \over \lambda -\mu} T^+(x, x_0) \tilde L(x_0) T^-(x_0, -A) T^+(A, x) \non\\
&& {\mathbb V}^-(x, \lambda, \mu) = {t^{-1} \over \lambda - \mu} T^-(x, -A)T^+(A, x_0)\tilde L(x_0) T^-(x_0,x) \non\\
&& \tilde {\mathbb V}^+(x_0, \lambda, \mu) = {t^{-1}(\lambda) \over \lambda -\mu}  \tilde L(x_0) T^-(x_0, -A) T^+(A,x_0)\non\\
&& \tilde {\mathbb V}^-(x_0,\lambda, \mu) =  {t^{-1}(\lambda) \over \lambda -\mu} T^-(x_0, -A) T^+(A,x_0) \tilde L(x_0).
\ee
The next step is to expand the latter expressions. Notice that special care is taken by construction
for the defect point, where separate formulaes naturally emerge.
Substituting the ansatz for the monodromy matrices we can explicitly derive this expansion.
Explicit expressions for the first three orders are given below:
\be
{\mathbb V}(\mu,\ x) = \begin{pmatrix}
1 & 0\\
0 & 0 \end{pmatrix}\label{v1}
\ee
\be
&&{\mathbb V}^{-(2)}(\mu,\ x) = \begin{pmatrix}
   \mu & \bar \psi^-(x) \\
    \psi^-(x) & 0
\end{pmatrix} \   \cr
&&{\mathbb V}^{+(2)}(\mu,\ x) = \begin{pmatrix}
   \mu & \bar \psi^+(x) \\
    \psi^+(x) & 0
\end{pmatrix} \ \cr
&&\tilde {\mathbb V}^{-(2)}(\mu,\ x_0) =\begin{pmatrix}
   \mu & \bar \psi^+(x_0)+ \beta(x_0) \\
    \psi^-(x_0) & 0
\end{pmatrix} \ ,\cr
&& \tilde {\mathbb V}^{+(2)}(\mu,\ x_0)=\begin{pmatrix}
   \mu & \bar \psi^+(x_0) \\
    \gamma(x_0) +\psi^-(x_0) & 0
\end{pmatrix} \ . \label{v2}
\ee
\be
&&{\mathbb V}^{-(3)}(\mu,\ x) = \begin{pmatrix}
  \mu^2- \bar \psi^{-}(x)\psi^-(x) & \mu \bar \psi^-(x) +\bar \psi^{-'}(x) \\
    \mu \psi^-(x) -\psi^{-'}(x)& \bar \psi^-(x)\psi^-(x)
\end{pmatrix} \ \non\\
&&{\mathbb V}^{+(3)}(\mu,\ x) = \begin{pmatrix}
   \mu^2 - \bar \psi^{+}(x)\psi^+(x) & \mu \bar \psi^+(x) +\bar \psi^{+'}(x) \\
    \mu \psi^+(x) -\psi^{+'}(x)& \bar \psi^+(x)\psi^+(x)
\end{pmatrix} \ \non\\
&&\tilde {\mathbb V}^{-(3)}(\mu,\ x_0) = \begin{pmatrix}
   \mu^2 -\Big (\bar \psi^{+}(x_0)+\beta(x_0)\Big )\psi^-(x_0) & \mu \Big (\bar \psi^+(x_0) +\beta(x_0)\Big ) +{\mathfrak f}(x_0)\\
    \mu \psi^-(x_0)-\psi^{-'}(x_0) & \Big (\bar \psi^+(x_0) +\beta(x_0)\Big )\psi^-(x_0)
\end{pmatrix} \ \non\\
&&\tilde {\mathbb V}^{+(3)}(\mu,\ x_0) = \begin{pmatrix}
   \mu^2 -\bar \psi^{+}(x_0)(\psi^-(x_0)+\gamma(x_0)) & \mu \bar \psi^+(x_0) +\bar \psi^{+'}(x_0) \\
    \mu \Big (\psi^-(x) +\gamma(x_0)\Big )+{\mathfrak g}(x_0) & \bar \psi^+(x_0)\Big (\psi^{-}(x_0) + \gamma(x_0)\Big )
\end{pmatrix} \ \non\\ \label{v3}
\ee
where we define
\be
&&{\mathfrak f}(x_0)= \bar \psi^{+'}(x_0) -\alpha(x_0)\Big (\beta(x_0) +2\bar \psi^{+}(x_0)\Big ) \non\\
&&{\mathfrak g}(x_0)= -\psi^{-'}(x_0) -\alpha(x_0)\Big (\gamma(x_0) +2\psi^{-}(x_0)\Big ). \label{fg}
\ee
Due to continuity requirements at the points $x_0^+,\ x_0^-$ (see also a similar argument in \cite{avandoikou}) i.e.
\be
{\mathbb V}^{+(k)}(x^+_0) \to \tilde {\mathbb V}^{+(k)}(x_0),
~~~~~~{\mathbb V}^{-(k)}(x^-_0) \to \tilde {\mathbb V}^{-(k)}(x_0), ~~~~x_0^{\pm} \to x_0
\ee
we end up with the following sewing conditions $C_{\pm}^{(k)}$ associated to the defect point:
\be
C_{-}^{(1)}: ~~~~&& \bar\psi^-(x_0) - \bar\psi^+(x_0)-\beta(x_0) =0, \non\\
C_{+}^{(1)}: ~~~~&& \psi^+(x_0)- \psi^-(x_0)-\gamma(x_0) =0 \non\\
C_{-}^{(2)}: ~~~~&& \bar \psi^{-'}(x_0) - \bar \psi^{+'}(x_0)+\alpha(x_0)\beta(x_0) +2 \alpha(x_0)\bar \psi^{+}(x_0) =0\non\\
C_{+}^{(2)}: ~~~~&& \psi^{-'}(x_0) - \psi^{+'}(x_0)+\alpha(x_0)\gamma(x_0) +2 \alpha(x_0)\psi^{-}(x_0)=0\non\\
C_{-}^{(3)}: ~~~~&& \bar \psi^{-''}(x_0) - \bar \psi^{+''}(x_0) + 2 \bar \psi^{-}(x_0)^2 \psi^{-}(x_0)-
\bar \psi^{+}(x_0)^2 \psi^{+}(x_0) + 2 \alpha(x_0) \bar \psi^{+'}(x_0)\non\\
&& -2 \Big (\beta(x_0) \bar \psi^{+}(x_0) + \beta^2(x_0)\Big ) \psi^{-}(x_0) + \Big (\beta(x_0) \gamma(x_0) - 2 \alpha^2(x_0)\Big ) \bar \psi^{+}(x_0) - \beta(x_0) \alpha^2(x_0) = 0\non\\
C_{+}^{(3)}: ~~~~&& \psi^{-''}(x_0) -\psi^{+''}(x_0) + 2 \psi^{-}(x_0)^2 \bar\psi^{-}(x_0)-\psi^{+}(x_0)^2 \bar\psi^{+}(x_0) + 2 \alpha(x_0)\psi^{+'}(x_0) \non\\
&& -2 \Big (\gamma(x_0) \psi^{+}(x_0) + \gamma^2(x_0)\Big ) \psi^{-}(x_0)  + \Big (\beta(x_0) \gamma(x_0) - 2 \alpha^2(x_0)\Big ) \psi^{+}(x_0) - \gamma(x_0) \alpha^2(x_0) = 0. \non\\
\label{sew2}
\ee
Higher ($j$-th) sewing conditions involving jumps of higher $(j-1)$-th derivatives of the fields will arise from the
construction of time components of higher Lax pairs.

Step $4$ of the procedure now follows:  we need to explicitly check the compatibility of the sewing conditions
with the hierarchy of Hamiltonian evolutions i.e we show that generic relationships of the type (\ref{sew})
can be implemented consistently with the commuting time evolutions. We have in particular
checked that:
\be
 \Big \{ {\cal H}^{(1)},\ C_{\pm}^{(1)} \Big \}   &=& \pm C_{\pm}^{(1)}, ~~~~~~\Big \{ {\cal H}^{(1)},\ C_{\pm}^{(2)} \Big \}
 = \pm C_{\pm}^{(2)} \non\\
\Big \{ {\cal H}^{(2)},\ C_{\pm}^{(1)} \Big \}&=& \mp C^{(1)}_{\pm} \delta(0) + C_{\pm}^{(2)}\non\\
\Big \{ {\cal H}^{(2)},\ C_{\pm}^{(2)} \Big \}
 &=&  C_{\pm}^{(3)} \pm C_{\pm}^{(1)} \delta'(0), \non\\
\Big \{ {\cal H}^{(3)},\ C_{-}^{(1)} \Big \} &=&  C_{-}^{(3)}+ \delta(0) C_{-}^{(2)}
+ \delta^{'}(0)C_{-}^{(1)}  + 2\bar\psi^{+ 2}(x_0) C_{+}^{(1)} \non\\ &+& \Big (\bar\psi^{+}(x_0) +\bar\psi^{-}(x_0) + 2 \beta(x_0)\Big )\psi^{-}(x_0) C_{-}^{(1)} \non\\
\ee
In section 4 we shall formally prove the compatibility of the generic sewing constraints, emerging
from continuity conditions imposed on the time components of the Lax pairs, with the charges in involution.

Having defined both the local integrals of motion, and the corresponding Lax pairs we now extract the associated equations of
motion. Hamiltonian equations on the one hand
\be
&&\dot \psi^{\pm}(x, t)= \{{\cal H}^{(j)},\ \psi^{\pm}(x,t)\}, ~~~~\dot {\bar \psi}^{\pm}(x, t)= \{{\cal H}^{(j)},\ \bar \psi^{\pm}(x,t)\}  ~~~~x \neq x_0 \non\\
&&\dot {\mathrm e}(x_0, t) = \{{\cal H}^{(j)},\ {\mathrm e}(x_0, t) \}, ~~~~{\mathrm e} \in \{\alpha,\ \beta,\ \gamma\}
\ee
and the zero curvature conditions ({\ref{zero1}), (\ref{zerod}) on the other hand, give rise as
should rightly be expected to the same equations of motion.
We shall focus on the equations of motion emerging from the Hamiltonian ${\cal H}^{(3)}$
(and the Lax pair ${\mathbb U}^{\pm},\ {\mathbb V}^{\pm(3)}$). For the left and right bulk theories we obtain the familiar equations of motion from the NLS model
\be
\dot \psi^{\pm}(x,t)  = {\partial^2 \psi^{\pm}(x, t) \over \partial x^2 } - 2 |\psi^{\pm}(x, t)|^2 \psi^{\pm}(x, t)\non\\
\dot {\bar \psi}^{\pm}(x,t)  = {\partial^2 \bar \psi^{\pm}(x, t) \over \partial x^2 } - 2 |\psi^{\pm}(x, t)|^2 \bar \psi^{\pm}(x, t)
\ee
the dot denotes derivative with respect to time.
For the defect point
\be
\dot \alpha(x_0)&=& \gamma(x_0) \bar \psi^{+'}(x_0) + \beta(x_0) \psi^{-'}(x_0) - \alpha(x_0) \gamma(x_0) \bar \psi^+(x_0) + \alpha(x_0) \beta(x_0) \psi^-(x_0) \non\\
\dot \beta(x_0) &=& 2 \alpha^2(x_0) \bar \psi^+(x_0) -2 \alpha(x_0) \bar \psi^{+'}(x_0) +\alpha^2(x_0) \beta(x_0) -
\beta(x_0)\gamma(x_0) \bar \psi^+(x_0) \non\\ &-&
\beta^2(x_0) \psi^-(x_0) -2 \beta(x_0) \bar \psi^+(x_0) \psi^-(x_0) \non\\
\dot \gamma(x_0)&=&  -2 \alpha(x_0) \psi^{-'}(x_0) -2 \alpha^2(x_0) \psi^-(x_0) - \alpha^2(x_0)\gamma(x_0) + \gamma^2 \psi^-(x_0) \non\\ &+& \beta(x_0)\gamma(x_0) \psi^-(x_0) +2 \gamma(x_0) \bar \psi^+(x_0) \psi^-(x_0).
\ee
The fact that we end up to the same equations of motion from either the Hamiltonian or the Lax
pair description for all the points on the line further confirms the consistency of the whole process.
Indeed, identification of the equations of motion from these two procedures confirms that the sewing conditions represent a
guarantee that the time-like operators of the Lax pair are correctly defined by the Semenov-Tjan-Shanskii
expression (\ref{timecomp}) on the rhs and lhs of the defect point consistently with the hamiltonian evolutions, in other words the sewing conditions may be conjectured to
represent the necessary and sufficient consistency conditions for application of the Semenov-Tjan-Shanskii procedure
to inhomogeneous forms of monodromy matrices.

\section{Discrete NLS model: the continuum limit}

We shall first briefly recall in this section the main results reported in \cite{doikou-defect}. From the discrete
expressions of the local integrals of motion and the corresponding Lax pairs we shall derive a consistent continuum limit
which will reproduce the results of the previous section further confirming the validity of the proposed process. Let us first recall
that the bulk Lax operator is given by (see e.g. \cite{kundura}):
\be
L_{aj}(\lambda) &=& \lambda D_j + A_j \non\\&=&  \begin{pmatrix}
   \lambda +{\mathbb N}_j  & x_j \\
    -X_j & 1
\end{pmatrix} \ \label{lmatrix}
\ee where ${\mathbb N}_j = 1- x_j X_j$ and the fields $x,\ X$ are canonical:
\be
\Big\{x_i,\ X_j \Big \} = \delta_{ij}.
\ee

The defect Lax operator is basically the ${\mathfrak sl}_2$ one expressed as:
\be
\tilde L_{an} &=& \lambda + \tilde A_{an} \non\\
&=&  \lambda + \begin{pmatrix}
   \alpha_n & \beta_n \\
    \gamma_n & \delta_n
\end{pmatrix} \ ,
\ee
the index $n$ simply denotes the position of the defect on the one dimensional spin chain.
Note that the $\tilde L$ matrix is required to obey the same ultra-local Poisson bracket structure as the bulk
matrices $L$ (\ref{lmatrix}) so that integrability is ensured. For this reason the elements $\alpha,\ \beta,\ \gamma,\ \delta$
satisfy the following exchange relations:
\be
&& \Big\{\alpha_n,\ \beta_n\Big \}= \beta_n \non\\
&& \Big\{\alpha_n,\ \gamma_n\Big \} = -\gamma_n \non\\
&& \Big \{\beta_n,\ \gamma_n \Big \} = 2 \alpha_n.
\ee

Inserting the defect at the $n$-th site of the one dimensional lattice the corresponding monodromy matrix is expressed as:
\be
T_a(\lambda) = L_{aN}(\lambda) L_{aN-1}(\lambda) \ldots \tilde L_{an}(\lambda) \ldots L_{a1}(\lambda) .
\ee
The trace of the monodromy matrix --the transfer matrix $t(\lambda)$-- as customary provides
a family of Poisson commuting operators guaranteeing the integrability of the system.
The expressions of the discrete integrals of motion obtained in \cite{doikou-defect} from the expansion
of $\ln t(\lambda)$ in powers of ${1\over \lambda}$:
\be
H^{(1)} &=& \sum_{j\neq n} {\mathbb N}_j + \alpha_n \non\\
H^{(2)} &=&   - \sum_{j \neq n, n-1} x_{j+1} X_j-{1\over 2}\sum_{j\neq n}{\mathbb N}_j^2
-x_{n+1}X_{n-1} - \beta_nX_{n-1} + \gamma_n x_{n+1} -{\alpha_n^2 \over 2}\non\\
H^{(3)} &=& -\sum_{j \neq n, n\pm 1} x_{j+1} X_{j-1} + \sum_{j \neq n, n-1} ({\mathbb N}_j + {\mathbb N}_{j+1})x_{j+1} X_j +
{1\over 3} \sum_{j\neq n} {\mathbb N}_j^3 + \tilde x_{n, n+1}{\mathbb N}_{n-1}  X_{n-1}\non\\
&+&  \tilde X_{n, n-1} x_{n+1}{\mathbb N}_{n+1} + \alpha_n\tilde x_{n, n+1} X_{n-1}  +\alpha_n \tilde X_{n, n-1} x_{n+1}-\tilde x_{n, n+1} X_{n-2} -x_{n+2} \tilde X_{n, n-1}+{{\alpha_n}^3\over 3} \non\\
\ee
where we define
\be
\tilde x_{n, n+1} &=& x_{n+1} + \beta_n
\non\\
\tilde X_{n, n-1} &=& X_{n-1}- \gamma_n.
\ee
Similarly the time component of the discrete Lax pairs $L,\ {\mathbb A}^{(j)}$
were explicitly derived in \cite{doikou-defect}, and the corresponding expressions are recalled below:
${\mathbb A}_j^{(1)}$ remains the same for all sites,
\be
{\mathbb A}_j^{(1)}(\mu) =  \begin{pmatrix}
    1 & 0 \\
    0 & 0
\end{pmatrix}
\ee
${\mathbb A}_j^{(2)}$ for $j \neq n,\ n+1$ is given by
\be
{\mathbb A}_j^{(2)}(\mu) =\begin{pmatrix}
   \mu & x_j \\
    -X_{j-1} & 0
\end{pmatrix},
\ee
 whereas
\be
{\mathbb A}_{n}^{(2)} = \begin{pmatrix}
   \mu & \beta_n + x_{n+1} \\
   -X_{n-1} & 0
\end{pmatrix}, ~~~~~{\mathbb A}^{(2)}_{n+1}= \begin{pmatrix}
   \mu & x_{n+1} \\
   \gamma_n -X_{n-1} & 0
\end{pmatrix}.
\ee
Also ${\mathbb A}_j^{(3)}$ for $j \neq n,\ n\pm 1,\ n +2 $ is given by
\be
{\mathbb A}_j^{(3)}= \begin{pmatrix}
   \mu^2  +x_j X_{j-1} & \mu x_j -x_j {\mathbb N}_j + x_{j+1}\\
    -\mu X_{j-1} + X_{j-1}{\mathbb N}_{j-1}-X_{j-2} & -x_jX_{j-1}
\end{pmatrix}
\ee
and
\be
{\mathbb A}_{n-1}^{(3)} &=& \begin{pmatrix}
\mu^2 + x_{n-1} X_{n-2}  &  \mu x_{n-1} +\tilde x_{n, n+1} -{\mathbb N}_{n-1} x_{n-1}\\
-\mu X_{n-2} -X_{n-3} +{\mathbb N}_{n-2} X_{n-2}
 & - X_{n-2}x_{n-1}
\end{pmatrix} \non\\
{\mathbb A}_{n}^{(3)} &=& \begin{pmatrix}
\mu^2 + \tilde x_{n, n+1} X_{n-1} &  \mu \tilde x_{n, n+1} +x_{n+1}
-{\mathbb N}_{n+1} x_{n+1}  +{\mathfrak f} \\
-\mu X_{n-1} -X_{n-2} + {\mathbb N}_{n-1} X_{n-1}  & -\tilde x_{n, n+1} X_{n-1}
\end{pmatrix} \non\\
{\mathbb A}_{n+1}^{(3)} &=& \begin{pmatrix}
\mu^2 + x_{n+1} \tilde X_{n,n-1}   &  \mu x_{n+1} +x_{n+2} -{\mathbb N}_{n+1} x_{n+1}\\
-\mu \tilde X_{n, n-1} -X_{n-1} +{\mathbb N}_{n-1} X_{n-1} +{\mathfrak g}
 &  - \tilde X_{n, n-1}x_{n+1}
\end{pmatrix} \non\\
{\mathbb A}_{n+2}^{(3)} &=& \begin{pmatrix}
\mu^2 + x_{n+2} X_{n+1}  &  \mu x_{n+2} +x_{n+3} -{\mathbb N}_{n+2} x_{n+2} \\
-\mu X_{n+1} -\tilde X_{n, n-1} +{\mathbb N}_{n+1} X_{n+1}
 & - X_{n+1}x_{n+2}
\end{pmatrix} \non\\
\ee
where we define
\be
{\mathfrak f} &=& x_{n+2} - x_{n+1} -\alpha_n (\beta_n +2 x_{n+1}) \non\\
{\mathfrak g} &=& X_{n-1} -X_{n-2} - \alpha_n(\gamma_n - 2 X_{n-1}). \label{fgc}
\ee
Notice that the continuum limits of the expressions above (\ref{fgc}) provide the continuum quantities given in (\ref{fg}).

To obtain the suitable continuum limits of the expressions defined above
let us first introduce the  spacing parameter $\Delta$ in the $L$-matrix of the discrete NLS
model as well as in the $\tilde L$ matrix of the defect (index-free notation):
\be
L(\lambda) =  \begin{pmatrix}
   1+ \Delta \lambda - \Delta^2 x X & \Delta x \\
    -\Delta X  & 1
\end{pmatrix} \
\ee
\be
\tilde L(\lambda) = \Delta \lambda + \Delta \begin{pmatrix}
   \alpha & \beta \\
    \gamma & \delta
\end{pmatrix} \
\ee

Let us now introduce the following notation (see also \cite{doikou-defect, ADS, doikoucl}):
\be
&&  x_j\ \to\ x^-(x), ~~~~~X_j\ \to\ X^-(x), ~~~~~1 \leq j \leq n-1, ~~~~~x\in (-A,\ x_0^-) \cr
&&  x_j \to x^+(x), ~~~~~ X_j  \to X^+(x), ~~~~~~n+1 \leq j \leq N, ~~~~~x \in (x_0^+,\ A),
\ee
where $x_0$ is the defect position in the continuum theory.
Note also that in order to perform the continuum limit we bear in mind that:
\be
&& \Delta\ \sum_{j=1}^{n-1} f_j \ \to\ \int_{-A}^{x_0^-}dx\ f^-(x) \non\\
&& \Delta\ \sum_{j=n+1}^{N} f_j \ \to\ \int_{x_0^+}^{A} dx\ f^+(x).
\ee
The continuum limit of the first integral of motion is then given by (\ref{h1}).

Notice that in the first integral we considered terms proportional to $\Delta$, whereas in the second integral
the first non trivial contribution to the continuum limit is of order $\Delta^2$. The respective continuum quantity reads
then as in (\ref{h2}).
The continuum limit of $H^{(3)}$, after taking into account terms of order $\Delta^3$, becomes (\ref{h3}).
It is clear that the expressions (\ref{h1})-(\ref{h3}) were obtained by simply identifying:
\be
x^{\pm} \equiv \bar \psi^{\pm}, ~~~~~X^{\pm} \equiv - \psi^{\pm}.
\ee
Moreover, in the continuum limit the Lax pair is formulated as:
\be
L_j(\lambda) \to {\mathbb I} + \Delta\ {\mathbb U}(\lambda, \ x) + {\cal O}(\Delta^2),
~~~~~{\mathbb A}_j \to {\mathbb V}(x), ~~~~~{\mathbb A}_{j+1} \to {\mathbb V}(x+\Delta). \label{contb}
\ee
The discrete zero curvature condition reads as:
\be
\dot L_j(\lambda) = {\mathbb A}_{j+1}(\lambda) L_j(\lambda) - L_j(\lambda){\mathbb A}_j(\lambda), ~~~~j\neq n
\ee
which in the continuum limit takes the form (keep terms of order $\Delta$) \cite{ADS, doikoucl}:
\be
\dot{\mathbb U} - {\mathbb V}' +\Big [{\mathbb U},\ {\mathbb V} \Big ] =0.
\ee
The Lax pair associated to the first integral is quite trivial (\ref{v1}).
The Lax pairs associated to the integrals of motion are derived
after taking the following limits:
\be
&& L_j \to L^+(x), ~~~~A_j^{(k)} \to {\mathbb V}^{+(k)}(x). ~~~~~~j \in \{ n+1,\ \ldots N  \}, ~~~~x\in (x_0^+,\ A) \non\\
&& L_j \to L^-(x), ~~~~A_j^{(k)} \to {\mathbb V}^{-(k)}(x). ~~~~~~j \in \{ 1,\ \ldots n-1\}, ~~~~x \in (-A,\ x_0^-) \non\\
&& \tilde L_n \to \tilde L(x_0), ~~~~A_n^{(k)} \to \tilde {\mathbb V}^{-(k)}(x_0), ~~~~~A_{n+1}^{(k)} \to \tilde {\mathbb V}^{+(k)}(x_0). \label{dict2}
\ee
Let us now comment on the zero curvature condition at the defect point. Recall the associated discrete zero curvature condition:
\be
\dot{\tilde L}_n(\lambda) = {\mathbb A}_{n+1}(\lambda) \tilde L_n(\lambda) - \tilde  L_n(\lambda) {\mathbb A}_n(\lambda).
\ee
The continuum limit of the latter formula, bearing also in mind (\ref{dict2}) is given by expression (\ref{zerod}).
The time component corresponding to ${\cal H}^{(2)}$ (terms of order $\Delta$) is then given by (\ref{v2})
And the quantities corresponding to ${\cal H}^{(3)}$ (terms of order $\Delta^2$) are given by (\ref{v3}).
The valid continuum limits taken above provide extra consistency checks on the results obtained in the continuum case in the previous section.

\section{Hamiltonian compatibility for the sewing conditions}

We shall formulate in this section a generic proof on the compatibility of the sewing conditions with the time evolutions triggered by the hierarchy of Hamiltonians.

A formal justification of the closure of sewing conditions on themselves under a linear evolution triggered
by the integrable Hamiltonians generated by $\ln\ t(\lambda)$ can actually be given once the linear time evolution of the time-like component of the Lax pair is established. It is convenient to start our proof in the frame of discrete integrable models and then consider the suitable continuum limit along the lines described in the previous section.

We recall that the discrete time evolution ${\mathbb A}_j$ is defined as follows: given the Lax matrix $L_j(\mu)$ its time evolution (discrete zero curvature condition) reads as:
\be
{\dot L}_j(\mu) = {\mathbb A}_{j+1}(\lambda,\ \mu) L_j(\mu) - L_j(\mu){\mathbb A}_j(\lambda,\ \mu). \label{n1}
\ee
The generating function of the local Hamiltonians (the trace of the monodromy matrix)
may be expanded as: $\ln t(\lambda) = \sum_i {H^{(i)} \over \lambda^i}$
and the generating function ${\mathbb A}$ reads as:
${\mathbb A}_j(\lambda,\ \mu) = \sum_{i} {{\mathbb A}_j^{(i)}(\mu) \over \lambda^i}$.

In the case where the $r$ matrix associated to the system is the Yangian solution the time operator may be expressed as  (see also \cite{doikoucl} and references therein for more details)
\be
{\mathbb A}_j(\lambda,\ \mu) = {t^{-1}(\lambda) \over \lambda -\mu} T(j-1, 1;\lambda) T(N,j;\lambda)
\ee
where we introduce the notation:
\be
T(i,j;\lambda)=L_i(\lambda)L_{i-1}(\lambda) \ldots L_j(\lambda) ,~~~~i>j. \label{n4}
\ee

Making use of the latter relations (\ref{n1})-(\ref{n4}) we deduce the time evolution of ${\mathbb A}_j$, (recall also $\{t(\lambda),\ t(\mu)\} =0$) i.e.:
\be
\Big \{\ln t(z),\ {\mathbb A}_j(\lambda, \mu) \Big \} = \Big [{\mathbb A}_j(z, \lambda),\ {\mathbb A}_j(\lambda, \mu) \Big ]. \label{basic}
\ee
This derivation was given here for simplicity in the discrete framework. It naturally extends, --especially given the continuum limit process described in the previous section (see also \cite{ADS, doikoucl})-- to the continuous case by: $j\ \to\ x$ ($x \equiv \Delta j$). Equation (\ref{basic}) is valid for all points including the defect point $n$. Similarly its continuum equivalent is valid at every point of the interval $(-A,\ A)$ including the defect point $x_0$.

Recall now that the sewing conditions are generated by the continuity condition relating ${\mathbb V}^{\pm}(x_0^{\pm})$ and $\tilde {\mathbb V}^{\pm}(x_0)$. From (\ref{basic}) it becomes possible to write the generic time evolution of the sewing conditions:
\be
&&\Big \{ \ln t(z),\ {\mathbb V}^{\pm}(x_0^{\pm},\lambda, \mu) - \tilde {\mathbb V}^{\pm}(x_0, \lambda, \mu) \Big \}= \non\\
&&= \Big [{\mathbb V}^{\pm}(x_0^{\pm},z, \lambda),\ {\mathbb V}^{\pm}(x_0^{\pm},\lambda, \mu) \Big] -
\Big [\tilde {\mathbb V}^{\pm}(x_0,z, \lambda),\ \tilde {\mathbb V}^{\pm}(x_0,\lambda, \mu) \Big] \non\\ &&= \Big [\Delta{\mathbb V}^{\pm}(z, \lambda),\ {\mathbb V}^{\pm}(x_0^{\pm},\lambda, \mu) \Big] +
\Big [\tilde {\mathbb V}^{\pm}(z, \lambda),\ \Delta {\mathbb V}^{\pm}(\lambda, \mu) \Big], \label{basic2}
\ee
where self-explanatorily $\Delta {\mathbb V}^{\pm} = {\mathbb V}^{\pm}(x_0^{\pm}) - \tilde {\mathbb V}^{\pm}(x_0)$. Sewing conditions are obtained from expansion of $\Delta {\mathbb V}^{\pm}(\lambda, \mu)$ in powers of $\lambda^{-1}$. The $m$-th power yields the $m$-th sewing condition.
Expansion of the relevant terms in (\ref{basic2}) yields:
\be
\ln t(z) = \sum_{k\geq 0} {{\cal H}^{(k)} \over z^k}, ~~~~~\Delta {\mathbb V}^{\pm}(\lambda, \mu)= \sum_{m\geq 0} {{\cal C}_{\pm}^{(m)}(\mu) \over \lambda^m}
\ee
Then equation (\ref{basic2}) becomes:
\be
&&\Big \{ \sum_{ k\geq 0} {{\cal H}^{(k)} \over z^k},\ \sum_{m \geq 0} {{\cal C}_{\pm}^{(m)}(\mu) \over \lambda^m}\Big \} = \non\\
&& = \Big [\sum_{k\geq 0} {{\cal C}_{\pm}^{(k)}(\lambda) \over z^k},\ \sum_{j \geq 0}{{\mathbb V}^{\pm(j)}(x_0^{\pm}, \mu) \over \lambda^j} \Big ] + \Big [\sum_{k \geq 0}{ \tilde {\mathbb V}^{\pm (k)}(x_0, \lambda) \over z^k},\ \sum_{j\geq 0} {{\cal C}_{\pm}^{(j)}(\mu) \over \lambda^j} \Big ]\label{sum1}
\ee

Further expanding quantities
${\cal C}_{\pm}^{(k)}(\lambda),\ \tilde {\mathbb V}^{\pm(j)}(\lambda) $ in powers of $\lambda$ as
\be
\tilde {\mathbb V}^{\pm(k)}(x_0, \lambda) = \sum_{i=0}^{k-1}\tilde {\mathbb V}^{\pm(k,i)}(x_0) \lambda^i, ~~~~~~
{\cal C}_{\pm}^{(k)}(\lambda)= \sum_{i=0}^{k-1}{\cal C}_{\pm}^{(k,i)} \ \lambda^i
\ee
and fixing $k$ and $m$ in (\ref{sum1}) leads to the following fundamental relation:
\be
\Big \{{\cal H}^{(k)} ,\ {\cal C}_{\pm}^{(m)}(\mu) \Big \}= \sum_{i=0}^{k-1} \Big [{\cal C}_{\pm}^{(k,i)},\ {\mathbb V}^{\pm(m+i)}(x_0^{\pm}, \mu) \Big ] + \sum_{i=0}^{k-1} \Big [\tilde {\mathbb V}^{\pm(k,i)}(x_0),\  {\cal C}_{\pm}^{(m+i)}(\mu) \Big ].
\ee
Now expanding  ${\cal C}_{\pm}^{(p)}(\mu),\ {\mathbb V}^{\pm(m-i)}(x_0^{\pm}, \mu)$ in powers of $\mu$:
\be
{\cal C}_{\pm}^{(p)}(\mu) = \sum_{l=0}^{p-1} {\cal C}^{(p, l)}_{\pm}  \mu^l, ~~~~~{\mathbb V}^{\pm(p)}(x_0^{\pm}, \mu) =\sum_{l=0}^{p-1} {\mathbb V}^{(p, l)}_{\pm}(x_0) \mu^l
\ee
and fixing $l$ in the sums above we conclude
\be
\Big \{{\cal H}^{(k)} ,\ {\cal C}_{\pm}^{(m,l)} \Big \}= \sum_{i=0}^{k-1} \Big [{\cal C}_{\pm}^{(k,i)},\ {\mathbb V}^{\pm(m+i,l)}(x_0^{\pm}) \Big ] + \sum_{i=0}^{k-1} \Big [\tilde {\mathbb V}^{\pm(k,i)}(x_0),\  {\cal C}_{\pm}^{(m+i, l)} \Big ].
\ee
${\cal C}_{\pm}^{(p,l)}$ are matrices with entries being the constraints of the type (\ref{sew2}) or linear combinations thereof.
The Poisson bracket of any Hamiltonian ${\cal H}^{(k)}$ with the generic scalar constraint $C_{\pm}^{(m, l)}$ is now expressed as finite  linear combination of the same scalar constraints,
and this concludes our formal proof on the Hamiltonian compatibility of the sewing conditions.

\section{Conclusions and perspectives}

Let us now summarize what we have achieved at this time. We have formulated a fully Hamiltonian framework for a description of a
Liouville-integrable point-like single defect on a continuous line in a bulk-integrable field theory. The defect is initially
introduced as a set of discrete dynamical variables independent of the bulk fields, constrained however by the requirement of having a
Poisson structure parametrized by the same classical $r$-matrix as the bulk space-like Lax operator. A hierarchy of Poisson-commuting
Hamiltonians is then derived canonically from the combined bulk-defect monodromy matrix. Sewing conditions are then imposed
by the requirement that the time-like operators of the Lax pair (describing each time evolution associated with each Hamiltonian
of the hierarchy) be defined consistently in the left and right neighborhood of the defect point. They fix relations between the
defect parameters, and left-right limits at the defect of successive derivatives of the bulk fields.

They were shown, first on the specific example of non-linear Schr\"{o}dinger equation, then following a general algebraic argument based
on the $r$-matrix structure, to be compatible with all time evolutions triggered by the hierarchy of Hamiltonians in the sense
that the Poisson bracket of any hamiltonian with any constraint closes (moreover linearly) on the ideal of functions
on the phase space generated by the constraints; hence they can be simultaneously imposed to the dynamical bulk and defect
variables evolving simultaneously under action of the full hierarchy. The hierarchy of Hamiltonians together with the hierarchy
of sewing constraints thus defines a Liouville-integrable system.

The sewing conditions, as already emphasized, can be understood as the consistency conditions for the existence of a canonical
Semenov-Tjan-Shanskii type construction for well-defined time-like operators in the Lax pair, i.e. operators
such that the zero-curvature conditions for the Lax connection give the same equations of motion as the Hamiltonian
evolution computed from the Poisson structure, both in the bulk and at the defect point.

Finally as a further consistency check the continuous dynamical equations were identified with a suitable scaling
limit of the discrete formulation in \cite{doikou-defect}.
This first allows to put into a clearer perspective a number of previous results on ``integrable'' defects. As already commented upon
the Lagrangian approach advocated in many works \cite{BCZ1}--\cite{cozanls} has difficulties in dealing with the fundamentally
Hamiltonian notion of Liouville integrability; it is in a sense a ``single-time'' approach instead of the multi-time approach
naturally associated with the notion of Hamiltonian hierarchy. ``Integrability'' in this framework essentially means that
the constructed conserved quantities are shown to be time-invariant under one single time-evolution (the one
associated to the first non trivial Hamiltonian of the hierarchy; usually the second or third one) which is of course
weaker than Liouville integrability. A formulation closer to ours can be found in \cite{newdefect}, where the B\"{a}cklund
transformation scheme of \cite{cozanls} is rewritten using a combined bulk-defect monodromy matrix similar to ours. A similar
formulation was proposed earlier by \cite{haku}. In both cases however the defect matrix is written directly as a function of the limit bulk variables (in our language, this means solving directly the sewing conditions to get an on-shell defect matrix). This
makes the analysis of the Poisson structure (required to speak of integrability) tricky since the on-shell defect
matrix should now have non-trivial PB's with the left and right bulk monodromy matrices- an issue which our ``off-shell
plus constraints'' approach eliminates. Note that by contrast, the Hamiltonian formulation of a point like B\"{a}cklund transformation by Sklyanin \cite{sklyaninbac} is precisely of this ``off shell'' type.

We can now comment on several possible future developments of our scheme:
An extension of the point-like single defect approach to other ultra-local integrable field theories should not
raise too many difficulties at least in its principle. Theories considered in \cite{BCZ1}--\cite{cozanls} are natural candidates
to this extension, which would then clarify the issue of actual Liouville integrability for the proposed defect models.
Multiple point-like defects will a priori be described by similar combinations of bulk and defect matrices with
independent defect parameters and sewing conditions at each defect point.
Extended defects should also be considered, in the spirit of the so-called Type II B\"{a}cklund transformation formalism
\cite{BCZ2, newdefect} for which one should work at providing a Hamiltonian formulation following the lines of our present
construction.

A very challenging question is raised when considering non ultra-local theories. The ultra-local form
of the Poisson structure in our example considerably simplifies the formulation of the time-like Lax operators
which are essential to our whole scheme. Non ultra-local PB's  are naturally \cite{maillet} associated not with
single $r$-matrices but with $r,\ s$ pairs parametrizing semiclassical reflection algebras. The monodromy
matrix structure is more complicated (general quadratic form with two matrices) hence the defect Poisson
structure will also need to be extended; in addition the issue of defect/bulk interaction through crossed Poisson brackets
(even off-shell!) must be addressed. Construction of the time-like Lax operators also becomes then a very non
trivial operation (see e.g. \cite{avandoikou}).

Finally, let us comment on possible approaches to quantum integrable defects
in the continuum. As indicated in the Introduction a construction based on
a $RTT$ quantum algebra was already proposed in \cite{haku, caudr, weston}, and is based on the construction
of a quantum monodromy matrix on a discrete lattice. Note that the monodromy matrix, and the
results presented in \cite{doikou-defect} are apparently valid in the quantum case as well.

An alternative approach also exists. It was developed by the Annecy
group in general cases \cite{annecydef1} and particularized
to NLS in \cite{annecydef2}. It uses an ab initio approach through construction
of factorizable scattering matrices realizing a ``reflection-transmission
algebra''.
It turns out that this approach admits a classical limit\footnote{We are
indebted to Eric Ragoucy for pointing out this fact to us.} and a  comparison with our
current results may be quite illuminating: indeed  it is not immediate
at this stage how classical sewing conditions gotten from a  Semenov-Tjan-Shanski
scheme of Lax pair construction may arise from this $RT$ algebra.
It is interesting to note that self-adjointness requirements on extensions of the quantum NLS Hamiltonian
[25] closely resemble the first two continuity conditions of the classical time-like Lax operator.

This opens a new, final avenue of investigation. It is possible to compare this classical limit of quantum transmission matrices with classical transmission amplitudes on the
classical defect. These may be obtained by explicitly solving the bulk-plus-defect NLS equation for soliton-like configurations through  application of classical direct/inverse scattering methods to the Lax  pair $L$ and ${\mathbb V}$ corresponding to the NLS Hamiltonian with the first two sewing conditions. We conjecture that the sewing conditions, being regularity conditions on
the operator describing the time evolution, will allow for the actual existence of non singular,
computable classical amplitudes for the soliton moving across to the defect, hence allowing for the
existence of (at least semi classical) transmission matrices.
\\
\\
{\bf Acknowledgments}\\
This work was supported by CNRS, Universit\'e de Cergy Pontoise, Patras
University and ANR Project DIADEMS (Programme Blanc ANR SIMI 1 2010-BLAN-0120-02).
J.A. wishes to thank Patras University Engineering Department, and A.D. thanks UCP and LPTM Cergy,
for their  mutual warm hospitality.

\end{document}